\documentclass[12pt,preprint]{aastex}

\newcommand{\text}[1]{\rm #1}

\slugcomment{}

\shorttitle{Gravitational Radiation from SASI in Core-Collapse Supernovae}
\shortauthors{Kotake et al.}

\begin{document}


\title{Gravitational Radiation from Standing Accretion Shock Instability in Core-Collapse Supernovae}

\author{Kei Kotake\altaffilmark{1}, Naofumi Ohnishi\altaffilmark{2}, and 
 Shoichi Yamada\altaffilmark{3,4}}
\affil{$^1$Division of Theoretical Astronomy, National Astronomical Observatory  Japan, 2-21-1, Osawa, Mitaka, Tokyo, 181-8588, Japan}
\email{kkotake@th.nao.ac.jp}
\affil{$^2$Department of Aerospace Engineering, Tohoku University,
6-6-01 Aramaki-Aza-Aoba, Aoba-ku, Sendai, 980-8579, Japan}
\affil{$^3$Science \& Engineering, Waseda University, 3-4-1 Okubo, Shinjuku,
Tokyo, 169-8555, Japan}
\affil{$^4$Advanced Research Institute for Science and Engineering, Waseda University, 3-4-1 Okubo, Shinjuku,
Tokyo, 169-8555, Japan}

\begin{abstract}
We present the results of numerical experiments, in which we study
how the asphericities induced by the growth of the standing accretion 
shock instability (SASI) produce the gravitational waveforms 
in the postbounce phase of core-collapse supernovae.
To obtain the neutrino-driven explosions, 
we parameterize the neutrino fluxes 
emitted from the central protoneutron star and approximate the neutrino 
transfer by a light-bulb scheme. We find that the waveforms due to the 
anisotropic neutrino emissions show the monotonic increase with time, 
whose amplitudes are up to two order-of-magnitudes larger than the ones 
from the convective matter motions outside the protoneutron stars. 
 We point out that the amplitudes begin to become larger 
when the growth of the SASI enters the nonlinear phase, 
in which the deformation of the shocks and the neutrino anisotropy become 
large.
 From the spectrum analysis of the waveforms, we find that the amplitudes 
from the neutrinos are dominant over the ones 
from the matter motions at the frequency below $\sim 100$ Hz, which 
 are suggested to be within the detection limits of the detectors in the next
 generation such as LCGT and the advanced LIGO for a supernova at 10 kpc.
 As a contribution to the gravitational wave background, we show that 
 the amplitudes from this source could be larger 
at the frequency above $\sim$ 1 Hz than the primordial gravitational wave 
backgrounds, but unfortunately, invisible to the proposed space-based 
detectors.
\end{abstract}

\section{Introduction}
The gravitational astronomy is now becoming a reality. In fact, the 
ground-based laser interferometers such as TAMA300 \citep{tama,tamanew} and 
the first LIGO \citep{firstligo,firstligonew} are beginning to 
take data at sensitivities where astrophysical
events are predicted. For the detectors including GEO600 and VIRGO, core-collapse supernovae
especially in our Galaxy, have been supposed to be the most plausible
sources of gravitational waves (see, for example, \citet{new,kotake_rev} 
for review). Since the gravitational wave (plus neutrinos) is the only tool 
which gives us the information in the innermost part of evolved massive stars, the detection is 
important not only for the direct confirmation of gravitational waves but
also for the understanding of supernova physics itself. 

So far, most of the theoretical predictions of gravitational waves
from supernovae have focused on the bounce signal in the 
context of rotational \citep{mm,ys,zweg,dimmel,fry,kotakegw,shibaseki,ott} 
and magnetorotational \citep{kotakegwmag,obergaulinger} core collapse. 
In most of the previous studies, the iron core prior to core-collapse  
 was assumed to rotate much more rapidly than predicted by the 
recent stellar evolution calculations \citep{heger05}. 
Recently, the initial rotation periods were estimated to be larger than 
$\sim$ 100 sec for the observed 
rotation periods of the radio pulsars 
\citep{ott_birth}. 
In such a slowly rotating case, the bounce signal becomes  too small 
to be detected even by the laser interferometers in the next generation 
 for a galactic supernova, owing to the suppression of the
rotation-induced deformation at core-bounce (see, e.g., \citet{kotakegwmag}).

Besides the rapid rotation of the cores, two other ingredients have been 
considered to be 
important in the
 much later phases after core bounce, namely convective motions and anisotropic
neutrino emissions. Both of them contribute to the non-spherical
parts in the energy momentum tensor of the Einstein equations, thus
 being the potential sources of the gravitational wave 
(see, \citet{kotake_rev} for a review). One of the possibility as the origin 
of the asphericities 
may be large scale density inhomogeneities formed in the central core prior 
to collapse 
(e.g., \citet{bazan,meakin}). \citet{fryersingle} performed 
three dimensional SPH 
simulations and pointed out that 
the neutrino-originated gravitational waves, 
which dominate over the one from the convections, are within the detection 
limits for the advanced LIGO for the galactic supernova (see also,
\citet{burohey,fryersingle,fryer04,muyan97}). 
Another possibility to induce
anisotropy is the (moderate) rotation of the core.
\citet{mueller} calculated the gravitational waves based on  
the two-dimensional (2D) Boltzmann transport simulations  
of slowly rotating core \citep{buras} and 
found that the neutrino-originated gravitational waves exceed 
 the bounce signal large enough to be detectable by the advanced LIGO 
with good signal-to-noise ratio for the galactic supernova (see, also 
\citet{kotake_rev} for the properties of 
neutrino-originated gravitational waves in the rapidly rotating case).
 More recently, the new ingredient of 
the gravitational-wave emissions is reported \citep{ott_new}, 
namely the g-mode excitations 
of the protoneutron stars, which was observed in the 2D approximate Boltzmann 
transport simulations at much later postbounce phase ($\sim 600$ ms) 
\citep{burr_new}. 

There is an another ingredient for producing large asphericity, 
to which much attention has been paid recently in the context of the  
studies about the explosion mechanisms, that is the so-called 
standing accretion shock instability (often called ``SASI'').
 In the numerical simulations by  \citet{blondin_03,scheck_04,blondin_05,ohnishi_1,ohnishi_2}, it was found that the standing shock wave is shown to be unstable to non-radial perturbations, and that the 
perturbations grow up to the non linear regime with clear low-mode ($\ell=1,2$)
dominance, leading to the global deformation
 of the shock wave later. Here $\ell$ stands for the azimuthal index of the Legendre polynomials. The importance of SASI is also stressed by the recent studies, demonstrating 
that such an explosion is favorable to reproduce the observed
synthesized elements of SN1987A \citep{kifo} and also to explain the 
origin of the natal kicks of young pulsars \citep{scheck_04}. 
These situations motivate 
us to study how the gravitational waveforms are originated from 
 the asphericities by SASI.

In this paper, we present the results of numerical experiments, in which we study how the asphericities induced by the growth of SASI
 produce the gravitational waveforms.
To obtain the neutrino-driven 
explosions, we parameterize the neutrino fluxes 
emitted from the central protoneutron star and approximate the neutrino 
transfer by the light-bulb scheme. Based on the long-term two dimensional 
hydrodynamic results, we calculate the gravitational waveforms.
By doing the spectrum analysis, we study the 
detectability of such signals from a nearby core-collapse supernova.
%
It is noted that much attention has been paid recently to the 
core-collapse supernovae as one of the promising sources of the 
cosmological gravitational wave backgrounds (see, e.g., \citet{buonanno} and 
references therein). 
Thus we calculate the SASI-induced 
gravitational wave backgrounds and discuss the 
detectability by the currently proposed space-based detectors such as 
LISA\footnote{see http://lisa.jpl.nasa.gov/}, BBO\footnote{see http://universe.nasa.gov/program/bbo.html}, and DECIGO
 \citep{seto}. 
 
The plan of this paper is as follows: In Section \ref{sec2}, we outline
  the initial models, the numerical methods, and shortly 
summarize the methods for calculating the waveforms.
 We show the main numerical results in Section \ref{sec3}.  We summarize and discuss our results in 
Section \ref{sec4}.  

\section{Numerical Methods and Models\label{sec2}}

\subsection{Initial Models and Hydrodynamics}
The numerical methods employed in this paper are essentially the same as 
those used in our previous paper \citep{ohnishi_1}. 
The basic evolution equations, describing the compressible accretion 
flows of matter attracted by the protoneutron star and 
irradiated by neutrinos emitted from the neutrino sphere, 
are written as follows,
\begin{equation}
 \frac{d\rho}{dt} + \rho \nabla \cdot \mbox{\boldmath$v$} = 0,
\end{equation}
\begin{equation}
 \rho \frac{d \mbox{\boldmath$v$}}{dt} = - \nabla P - \rho\nabla\Phi,
\end{equation}
\begin{equation}
 \rho \frac{d}{dt}\displaystyle{\Bigl(\frac{e}{\rho}\Bigr)}
  = - P \nabla \cdot \mbox{\boldmath$v$} + Q_{\text{E}}
  + Q_{\rm inel} - Q_{\nu\bar{\nu}} - Q_{\rm brems} - Q_{\rm plasmon},
  \label{eq:energy}
\end{equation}
\begin{equation}
 \frac{dY_{\rm e}}{dt} = Q_{\text{N}},
  \label{eq:ye_flow}
\end{equation}
\begin{equation}
 \Phi = - \frac{G M_{\rm in}}{r},
  \label{eq:domain_g}
\end{equation}
where $\rho, \mbox{\boldmath$v$}, e, P, Y_{\rm e}, \Phi$ are
density, velocity, internal energy, pressure, electron fraction,
and gravitational potential, respectively.
We denote the Lagrangian derivative as $d/dt$ and $r$ is the radius.
 $M_{\rm in}$ is the mass of the central object.
The self-gravity of matter in the accretion flow is ignored 
(see \citet{yamasaki_06} for the effect).
The parameters of $Q_{\rm E}$ and $Q_{\rm N}$ are related to the 
standard heating and cooling via neutrino absorptions and emissions by free 
nucleons (see also \citet{ohnishi_1}). $Q_{\rm inel}$ is the minor additional heating by the inelastic 
 neutrino-helium interactions estimated by \citet{haxton} as considered in 
 \citet{ohnishi_2}.
 In addition, we newly take into account 
the neutrino cooling by neutrino 
pair annihilation to $e^{-}e^{+}$ pairs (: $Q_{\nu\bar{\nu}}$), nucleons-nucleons 
 bremsstrahlung (: $Q_{\rm brems}$) \citep{hannestad,burrows00}, and 
plasmon decay into $\nu_e, \bar{\nu}_{e}$ (: 
$Q_{\rm plasmon}$) \citep{schinder,ruffert}. 

The numerical code employed
in this paper is based on the modified version of the ZEUS-2D \citep{stone}
 for the applications to the supernova studies \citep{kotake}, in which 
 the tabulated realistic equation of state (EOS) 
based on the relativistic mean field theory \citep{shen98} was implemented. 
Furthermore, we have added the equation for electron fraction
(Eq.~(\ref{eq:ye_flow})) and included the neutrino coolings/heatings 
parametrically as the source term of the energy equations (Eq. (\ref{eq:energy})), both of which are solved in the operator-splitting fashion.
In the simulations, spherical coordinates are used without imposing the 
equatorial symmetry. The computation domain covers
the whole meridian section with 60 angular mesh points,
except for a model in which we have adopted 120 angular grid points. 
Since the latter model did not produce any significant difference
from other models, 
we will report in the following the results obtained
from the models with 60 angular mesh points.
We use 300 radial mesh points to cover
$r_{\rm in} \leq r \leq r_{\rm out} = {2000} ~{\rm km}$,
where $r_{\rm in}$ is the inner boundary and
chosen to be roughly the radius of neutrino sphere.

The initial conditions are provided in the same manner of
\citet{ohnishi_1} as the steady state solution of
\citet{yamasaki}. In constructing the initial conditions, we assume
 a fixed density $\rho_{\rm in} = 10^{11}$~g~cm$^{-3}$ at the inner boundary.
And the initial mass accretion rates and the initial 
mass of the central object are set 
to be $\dot{M} = 1~M_{\odot}$~s$^{-1}$ and $M_{\rm in} = 1.4~M_{\odot}$,
respectively.
To induce the non-spherical instability, we have added 
$\ell = 1$ velocity perturbations to the initial state mentioned above.
 At the outer boundary, we adopt the fixed boundary condition consistent 
with the initial condition. On the other hand, the absorbing boundary 
is used at the inner boundary.
 The temperatures of electron-type neutrinos are also 
constant and set to be $T_{\nu_{\text{e}}} = 4$~MeV and
 $T_{\bar{\nu}_{\text{e}}} = 5$~MeV,
which are the typical values in the post-bounce phase. 
The temperature of mu and tau neutrinos is 
chosen to be $T_{\nu_{\mu}} = 10$~MeV. In the standard model,
the luminosity of electron-type neutrino $L_{\nu_{\rm e}}$
and anti-neutrino $L_{\bar{\nu}_{\rm e}}$ are
set to be $6.5\times 10^{52}$~ergs~s$^{-1}$. In addition, we examined two 
cases of lower luminosities of $L_{{\nu}_{\rm e}}=L_{\bar{\nu}_{\rm e}} = 6.0, 5.5 \times 
10^{52}$~ergs~s$^{-1}$.
 In all the computed models, the luminosity of mu and tau neutrinos is set to be half value of electron-type 
neutrinos, keeping consistency with the results obtained by
the previous detailed numerical studies (e.g., \citet{lieb01}).

\subsection{Computations of gravitational wave signals}
\begin{figure}[hbt]
\epsscale{0.5}
\plotone{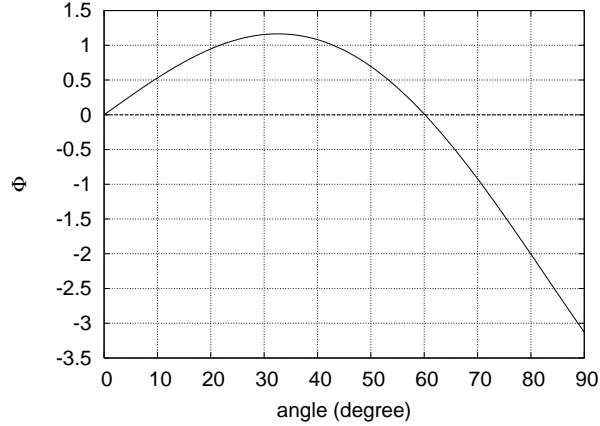}
\caption{Angular dependence of $\Phi$ in Eq. (\ref{graph1}). Note that the angle is measured
 from the symmetry axis.}
\label{phi}
\end{figure}

 We follow the methods based on \cite{epstein,muyan97,burohey,kotake_rev} in
order to compute the gravitational waveform from anisotropic neutrino 
emissions. We will summarize it in the following for convenience.
  Since we assume axisymmetry, the transverse-traceless gravitational field
$h^{\rm TT}$ is shown to have one nonvanishing component for the observer 
in the
equatorial plane:
\begin{equation}
h^{\rm TT}_{\nu} = \frac{4G}{c^4 R} \int_{0}^{t} dt^{'}
\int_{0}^{\pi}~d\theta^{'}~\Phi(\theta^{'})~\frac{dl_{\nu}(\theta^{'},t^{'})}{d\Omega^{'}},
\label{tt}
\end{equation}
where $G$ is the gravitational constant, $c$ is the speed of light, $R$
is the distance of the source to the observer 
, $dl_{\nu}/d\Omega$
represents the direction-dependent neutrino luminosity emitted per unit
of solid angle into direction of $\Omega$,  and $\Phi(\theta^{'})$ denotes
the quantity, which depends on the angle measured from the symmetry axis 
($\theta^{'}$) (see Figure \ref{phi}),
\begin{eqnarray}
\Phi({\theta^{'}})&=& \sin \theta^{'} 
\Bigl(-  \pi +  \int_{0}^{2\pi} d\phi^{'}
\frac{1 + \sin\theta^{'}\cos\phi^{'}}{1 + \tan^2\theta^{'}\sin^2\phi^{'}}\Bigr)
\\ &=&  \pi \sin \theta^{'} ( - 1 + 2 | \cos  \theta^{'}| ).
\label{graph1}
\end{eqnarray} 
 The integration with respect to the
azimuthal angle ($\phi^{'}$), albeit fairly straightforward, 
is pretty helpful to the later discussions, which is not published to our 
knowledge. In addition, from Eq. 
(\ref{graph1}), one can readily see that no
gravitational waves are emitted if the neutrino radiation is isotropic. 
We estimate $dl_{\nu}/d\Omega$ as follows,
\begin{equation}
\frac{dl_{\nu}}{d\Omega}(\theta^{'}) = \int_{0}^{r_{\rm out}}  dr~r^2~|Q_{\rm C}|
\Big|_{{\rm calculated~at~each}~\theta^{'}},
\end{equation}
where $Q_{\rm C}$ is the sum of contributions from the neutrino coolings, 
of each species ($\nu_{\rm e}$, $\bar{\nu}_{\rm e}$, $\nu_{\mu,\tau}$, and 
$\bar{\nu}_{\mu,\tau}$) computed from the source term
of Eq.~(\ref{eq:energy}). In the above estimation, neutrinos are assumed to be emitted radially in each angular bin. 
As a guide to see the anisotropy of the neutrino emissions, 
we calculate the anisotropy parameter according to \citet{muyan97} as follows,
\begin{equation}
\alpha(t) = \frac{1}{l_{\nu}(t)}\int_{0}^{\pi}~d\theta^{'}~\Phi(\theta^{'})~\frac{dl_{\nu}(\theta^{'},t)}{d\Omega^{'}}.
\label{anisop}
\end{equation}
It should be noted that the gravitational waves from neutrinos have a 
different feature from the ones from matter motions, in the sense that 
the former has {\it memory effect}, which means that the 
gravitational amplitude jumps from zero to a nonvanishing value and 
it keeps the non-vanishing value even after the energy 
source of gravitational waves disappeared (see \citet{braginsky} for details, 
and \citet{segalis,hiramatsu} for the examples of the 
astrophysical emitters of such gravitational waves and references therein). 
In Eq. (\ref{tt}), this nature can be directly seen as the time-integral.

As for the gravitational waves, $h(t)$, of the quadrupole radiation of mass
motions, we employ the standard strain formula,
\begin{equation}
h(t) = \frac{1}{8}\Biggl(\frac{15}{\pi}\Biggl)^{1/2} {\sin}^2 \alpha \,\,\frac{A_{20}^{\rm{E} 2}} {R},
\label{htt}
\end{equation}
where the form of $A_{20}^{\rm{E} 2}$ can be found in references (see 
equation (12) in \cite{mm}). In using the formula, the azimuthal
gradient of the gravitational potential is set to be zero, since 
we neglect the self-gravity of the computational domain and treat the 
gravity as in Eq. (\ref{eq:domain_g}).
In the following computations, we assume that the
observer is located in the equatorial plane since the 
 most of gravitational wave is radiated in the plane ($\alpha = \pi/2$ in
Eq. (\ref{htt})), and that the source is assumed to be located at our galactic
center ($R = 10~\rm{kpc}$).
In order to assess the detectability of the gravitational waves obtained
in this study, we employ the characteristic gravitational wave strain,
\begin{equation}
h_{\rm c}(f) = \frac{1}{R}\sqrt{\frac{2}{\pi^2}\frac{G}{c^3}
\frac{d E_{\rm GW}(f)}{d f}},
\label{fourier1}
\end{equation}
for a given frequency $f$ \citep{flanagan}. Here ${d E_{\rm GW} (f)}/{d f}$ is the energy
spectra of gravitational waves defined as follows,
\begin{equation}
\frac{d E_{\rm GW}(f)}{d f} = \frac{c^3}{G}\frac{(2\pi f)^2}{16 \pi}
\Bigl|\tilde{A}^{E2}_{20}(f)\Bigr|^2
\label{fourier2}
\end{equation}
We employ the Fast Fourier Transformation technique in order to
perform the Fourier transformation of $A^{E2}_{20}(t)$ to  $\tilde{A}^{E2}_{20}(f)$. 

We discuss the contribution of the 
gravitational waves mentioned above to the cosmological 
background gravitational radiation. 
For simplicity, we assume that all core-collapse supernovae have identical 
 emission characteristics. According to \citet{phinney,buonanno}, the 
density parameter, which represents the sum 
of energy densities radiated by a large number of independent core-collapse
 supernovae at each redshift, can be written as,
\begin{equation}
\label{los}
\Omega_{\rm gw}(f)=\frac{1}{\rho_{\rm c}c^2}\int_0^\infty dz\,
\frac{R_{\rm SN}(z)}{1+z}\left|\frac{dt}{dz}\right|
f_z\frac{dE_{\rm GW}}{df}(f_z)\,,
\end{equation}
where $\rho_{\rm c} =
3H_0^2/(8 \pi G)$ is the cosmic critical density, $R_{\rm SN}(z)$ 
is the event rate of core-collapse supernovae 
per comoving volume, and 
$f_z\equiv(1+z)f$. The cosmological model enters with
$|dt/dz|=[(1+z)H(z)]^{-1}$ and, for a flat geometry,
\begin{equation}
\label{cosmo}
H(z)= H_0\left[\Omega_{\rm M}(1+z)^3+\Omega_{\Lambda}\right]^{1/2}\,.
\end{equation}
We will use the parameters $\Omega_{\rm M}=0.3$,
$\Omega_{\Lambda}=0.7$, and $H_0=h_0\,100~{\rm km}~{\rm s}^{-1}~{\rm
Mpc}^{-1}$ with $h_0=0.72$.
As for the core-collapse supernova rate, we employ the parameterization 
by \citet{buonanno} as follows,
\begin{equation}
\label{evol}
R_{\rm SN}(z)=R_{\rm SN}^0\times\left\{\begin{array}{ll}
(1+z)^\beta & \mbox{for $z<1$}\\
2^{\beta-\alpha}(1+z)^\alpha &
\mbox{for $1\leq z\leq20$}\end{array}\right..
\end{equation}

We take $\beta=2.7$ and the
present-day rate $R_{\rm SN}^0=2\times 10^{-4}\,{\rm Mpc}^{-3}\, {\rm
yr}^{-1}$, which are consistent with the Super-K limits on the
diffuse neutrino background \citep{malek}. Since the parameter
$0\lesssim\alpha\lesssim2$ is much less constrained than $\beta$ and
$R_{\rm SN}^0$ (see, for example, \citet{ando_sato} and references therein), 
we examine the two cases of $\alpha =0, 2$ in the later discussion.
\begin{deluxetable}{ccccc}
\tabletypesize{\scriptsize}
\tablecaption{Model Summary \label{table1}}
\tablewidth{0pt}
\tablehead{
\colhead{$L_{\nu}$ ($10^{52}$ erg/s)}
 & \colhead{$\Delta t~~({\rm ms})$}
 & \colhead{$E_{\rm GW}$ ($M_{\odot} c^2$)}
 & \colhead{$f_{\rm eq}~~({\rm Hz})$}
 & \colhead{$h_{\rm eq}$}}
\startdata
 6.5 & 540    & $6.1\times 10^{-10}$ & 51.0  & $5.3\times 10^{-22}$ \\
 6.0 & 1000   &  $4.6 \times 10^{-10}$ & 59.0  & $7.0\times 10^{-22}$ \\ 
 5.5 & 1000   & $2.3\times10^{-10}$ &63.2  & $5.1\times 10^{-22}$ \\
\enddata
\tablecomments{%
$L_{\nu}$ denotes the input luminosity.
$\Delta t$ represents the simulation time.  
$E_{\rm GW}$ is the total radiated energy in the form of gravitational 
waves in unit of $M_{\odot} c^2$.
$f_{\rm eq}$ and $h_{\rm eq}$ represents the frequency,
 and amplitude, under which the gravitational 
waves from anisotropic neutrino emissions 
dominate the ones from the matter motions, respectively. 
Note that the supernova is assumed to be located at the distance of 10 kpc.}
\end{deluxetable}

\section{Results\label{sec3}}
Only in the case of 
$L_{\nu_e} = 6.5 \times 10^{52}~{\rm erg}~{\rm s}^{-1}$, we can observe the 
continuous increase of the average shock radius, reaching the outer 
boundary of the computational domain (2000 km in radius) in $\sim 500$ ms 
with the explosion energy of $4.0 \times 10^{50}$ ergs.
In other models, we terminate the simulations at about 1 sec, not seeing 
the increase of the shock radius. Bearing in mind the evidences that 
the SASI-induced explosions are favorable for explaining the observed 
quantities of supernovae mentioned earlier, we take the exploding model 
of $L_{\nu} = 6.5 \times 10^{52}~{\rm erg}~{\rm s}^{-1}$ as a reference
in the following. For later convenience,
 the values of several important quantities are summarized in Table
 \ref{table1}.
\begin{figure}[hbt]
\epsscale{1.2}
\plottwo{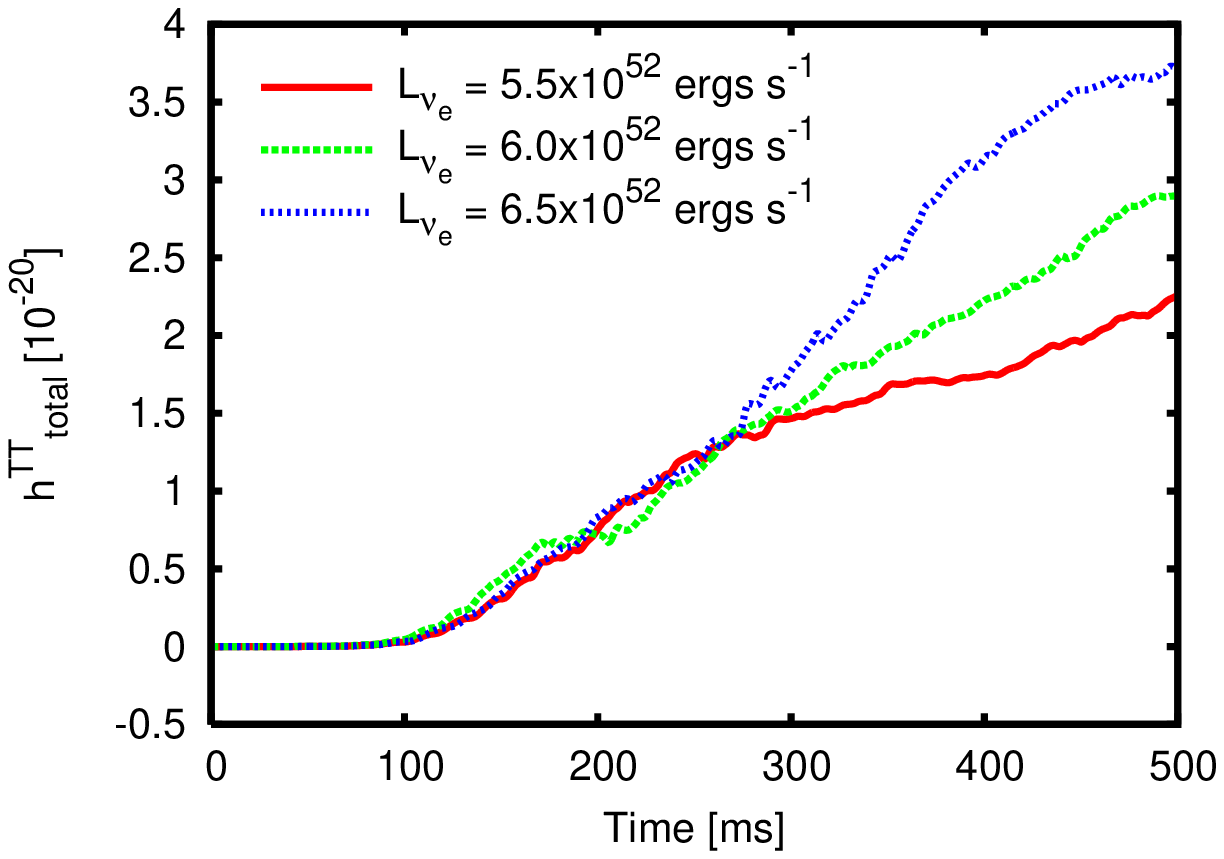}{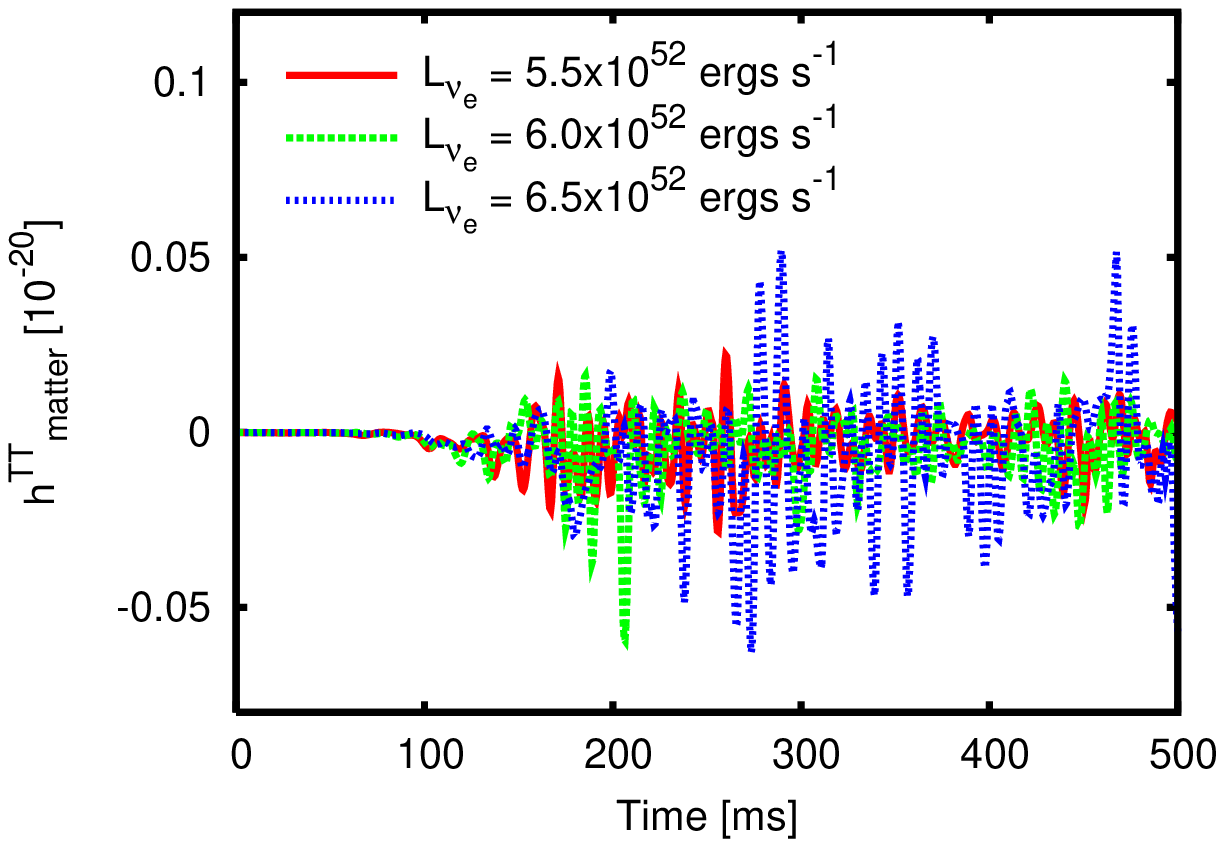}
\caption{Gravitational waveforms from the 
sum of the anisotropic neutrino emissions and the matter motions (left panel)
 and only from the matter motions (right panel). The time is measured from 
the epoch when the neutrino luminosity is injected from the surface of the 
neutrino sphere.
Note that the supernova is assumed to be located at the distance of 10 kpc.}
\label{fig2}
\end{figure}
\subsection{Properties of Waveform}
 In the left panel of Figure \ref{fig2}, the total amplitudes including 
the contributions both from the anisotropic neutrino emissions and
the matter motions,
are shown. Comparing the right panel, in which the amplitudes only from the 
matter motions are shown, we can see the neutrino-originated 
gravitational waves are up to two order-of-magnitudes larger than those  
 from the matter motions during the SASI operation (at $t \gtrsim 200$ ms as 
explained shortly). 
This reflects the small mass in the regions outside the neutrino sphere 
in the postbounce phase (at most $0.1 M_{\odot}$ throughout 
the simulation time). 
In the following, we pay attention to the neutrino-originated gravitational 
wave.

 Looking at the left panel of Figure \ref{fig2}, the waveforms 
show the monotonic increase with time regardless of the input 
neutrino luminosities. 
To understand this trend, we look into the hydrodynamical behaviors 
induced by the SASI, taking the case of $L_{\nu_e} = 6.5 \times 10^{52}~{\rm erg}~{\rm s}^{-1}$.

In Figure \ref{color1}, snapshots showing the hydrodynamical 
features with the resulting gravitational wave amplitudes (inserted figures) 
 are shown. Up to $\sim$ 100 ms after the onset of the simulation, 
no significant changes in the amplitudes and no deviations of the 
dynamics from spherical symmetry  are found.
It is noted that no gravitational waves are emitted when 
the motion of the regions outside is spherical, since the neutrino luminosity 
from the center is taken to be isotropic. 

\begin{figure}[hbt]
\epsscale{0.5}
\plotone{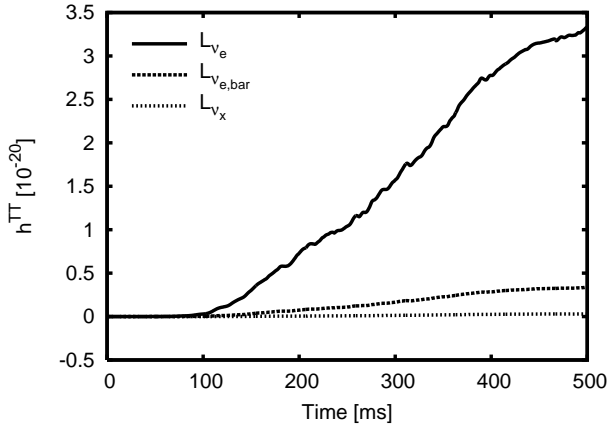}
\caption{Contribution of each neutrino species to the waveform for the case of 
$L_{\nu_e} = 6.5 \times 10^{52}~{\rm erg}~{\rm s}^{-1}$.
Here $\nu_{X} $ represents $\mu,\tau$ neutrinos. It is noted that
 the qualitative features are common to the other cases.}
\label{contribution}
\end{figure}

\begin{figure}
\epsscale{0.8}
\plotone{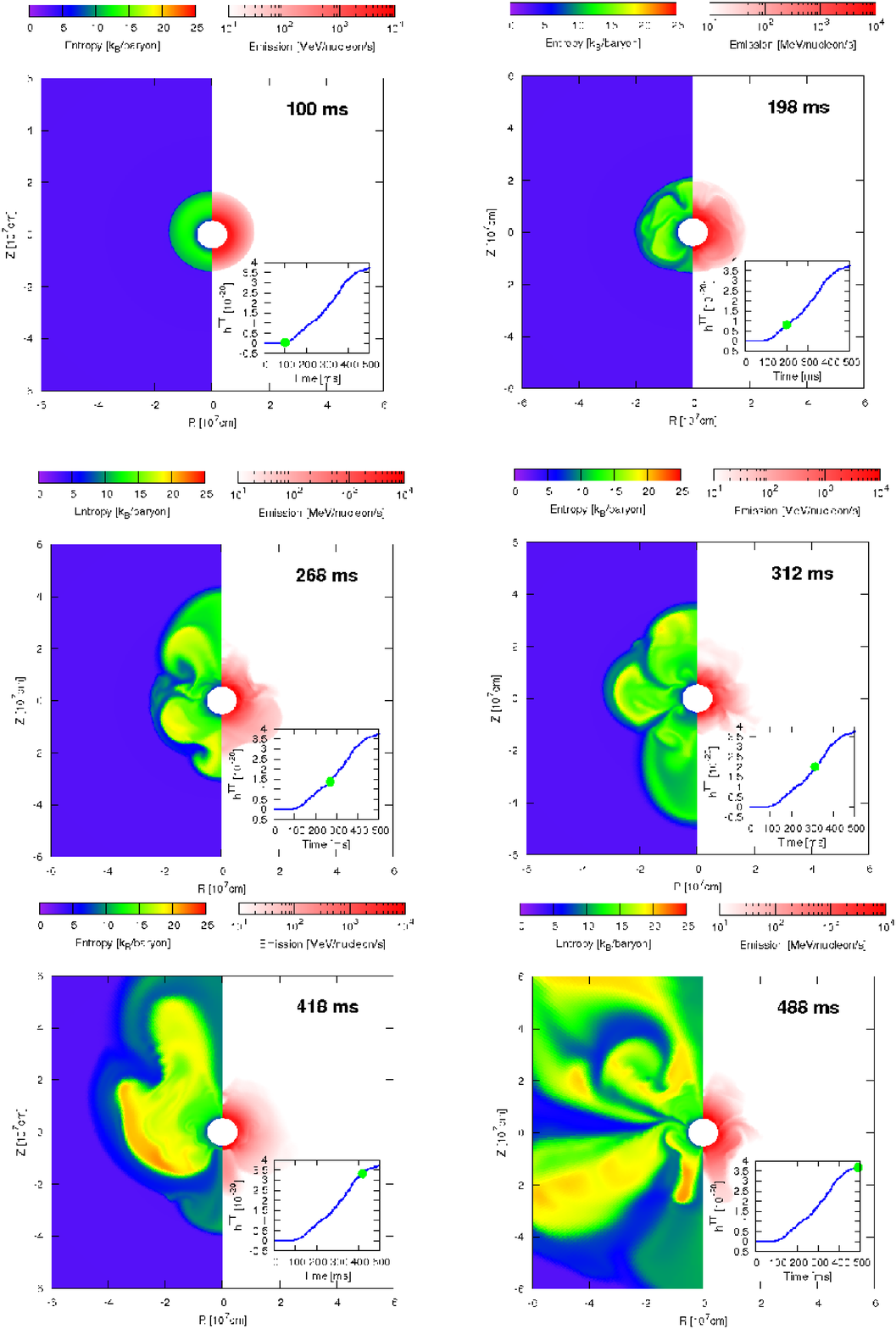}
\caption{Entropy- (the left half panel of each panel) and net neutrino 
emissivity- (the right half) distributions in the meridian section for the 
case of $L_{\nu_e} = 6.5 \times 10^{52}~{\rm erg}~{s}^{-1}$. The insert of 
each panel shows the wave amplitudes, in which the green point indicates 
the time of the snapshot.}
\label{color1}
\end{figure}

After $\sim$ 100 ms, the regions outside the neutrino sphere begin 
to oscillate with increasing average 
radius and wave amplitudes.
The reason why the sign of the growing amplitudes is positive 
 is as follows.
 From the angular dependence of $\Phi$ in Figure \ref{phi} and Eq. 
(\ref{graph1}), it can be seen 
that the neutrino emissions from the regions with $\theta^{'}$ 
smaller than $\pi/3$ contribute to the positive
 amplitude, while the negative sign comes from the regions from 
$\pi/3$ to $\pi/2$. 
It is noted that this feature is north-south symmetric (
Eq. (\ref{graph1})). Due to the dominance of $l=1,2$ modes of the 
deformed shock waves during the SASI operation mentioned below, 
the neutrino emissions become stronger
 in the regions close to the symmetry axis. 
 By these two factors, 
 the amplitudes are found to increase monotonically with time. 
This property is common to the lower luminosity models as seen from 
 the left panel of Figure \ref{fig2}. As a side-remark, the contribution
 of each neutrino species to the waveform is presented in Figure 
\ref{contribution}, which also shows the dominant contribution from $\nu_e$.

\begin{figure}[hbt]
\epsscale{1.1}
\plottwo{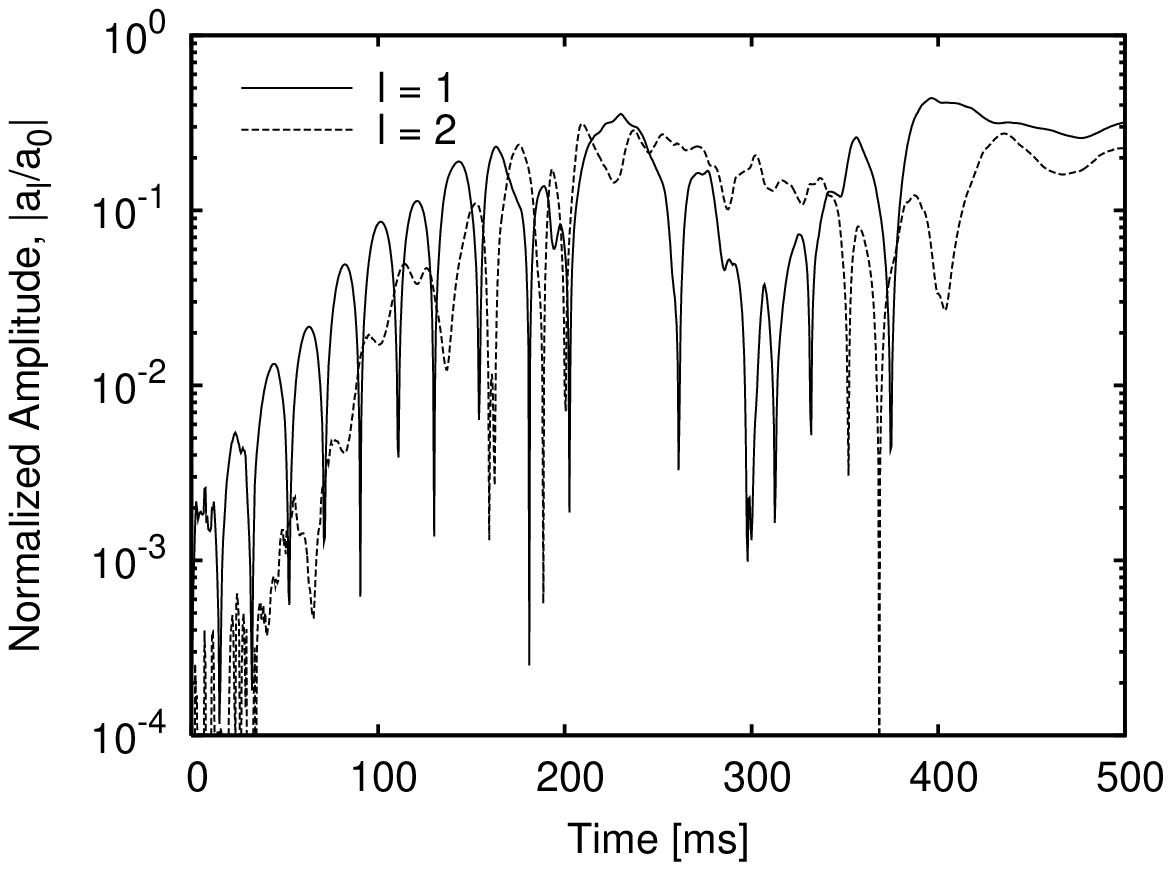}{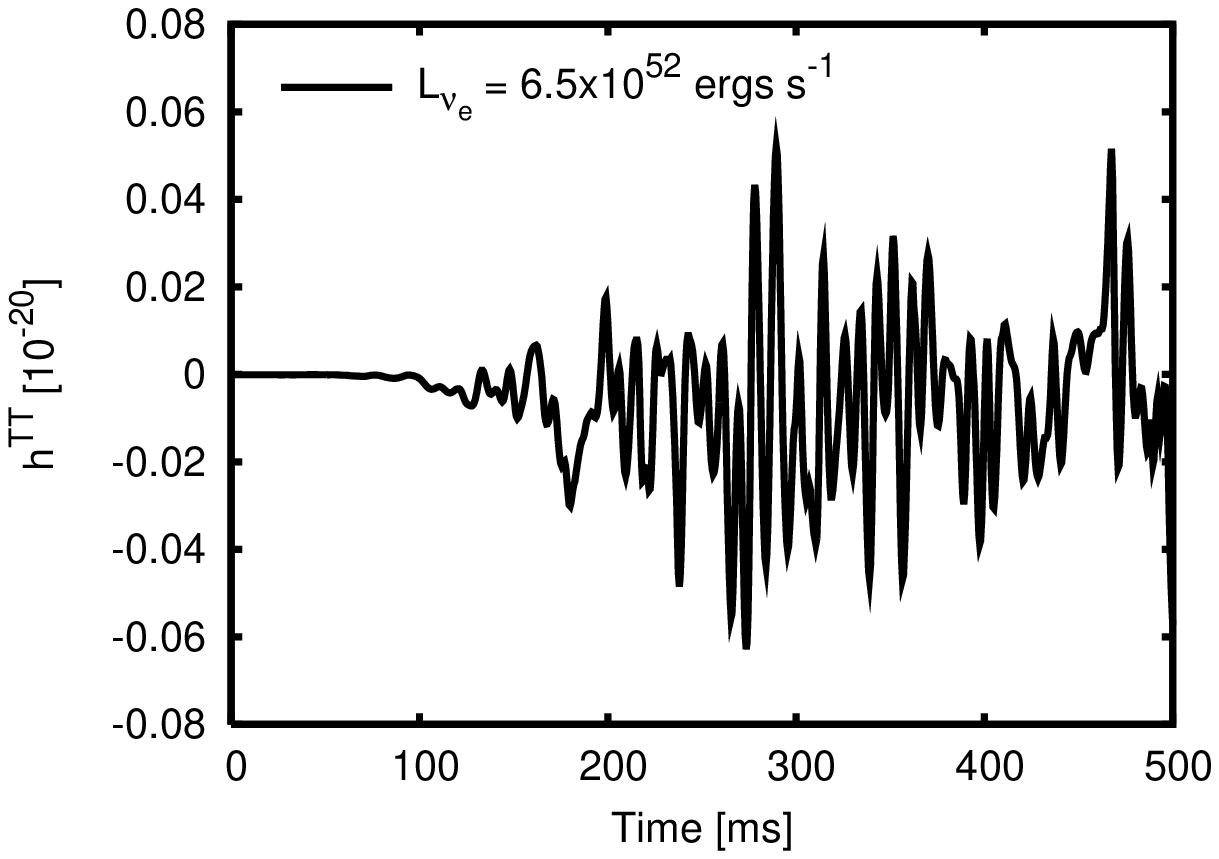}
\caption{Time evolution of the normalized amplitudes of $\ell = 1,2 $ modes 
 (left) and the waveform from the matter motions (right) for the case of 
$L_{\nu_e} = 6.5 \times 10^{52}~{\rm erg}~{\rm s}^{-1}$.}
\label{saturation}
\end{figure}

Next we discuss in more detail 
how the features of the waveforms are related to the growth of the SASI.
For the purpose, we decompose the fluctuations of 
the shock surface into the spherical harmonic components;
\begin{equation}
 R_{\rm s}(\theta) = \sum_{\ell=0}^{\infty}a_{\ell}
  \sqrt{\frac{2\ell+1}{4\pi}}P_{\ell}(\cos\theta).
\end{equation}
Since the system is axisymmetric, only $m=0$ harmonics, that is Legendre 
polynomials, show up.  The coefficients, $a_{\ell}$, can be calculated by 
the orthogonality of the Legendre polynomials;
\begin{equation}
 a_{\ell} = \frac{2\ell+1}{2}\int_{-1}^{1}R_{\rm s}(\theta)
  P_{\ell}(\cos\theta)d\cos\theta.
\end{equation}
The position of the shock surface, $R_{\rm s}(\theta)$, is estimated from
the iso-entropic surface of $s=5$. 
The left panel of Figure \ref{saturation} shows the evolution of 
the amplitude of each mode ($\ell = 1,2$) normalized by the average 
shock radius of $a_0$. The dominance of the fundamental mode ($\ell=1$) 
initially is simply due to the initial velocity perturbation assumed 
in this simulation.  
From $\sim 40~\mbox{ms}$, $\ell = 2$ begins to develop rapidly and at 
$\sim 200~\mbox{ms}$, the amplitude becomes of the same order as that of
 the fundamental 
mode, which has already been saturated by this time. 
This marks the beginning of the nonlinear phase. 

This transition from linear to 
nonlinear phase corresponds to the time of the rapid increase of the 
average shock radius as seen from the top right ($\sim 200$ ms) to middle 
left panel ($\sim 270$ ms) in Figure \ref{color1}. Simultaneously, 
 this makes the deformation of the shock more elongated 
 along the symmetry axis, leading to the slightly steep rise of the 
neutrino-originated gravitational wave afterwards for the 
reason mentioned above (see the left panel of Figure \ref{fig2}).  
Here it should be noted that the dominance of $\ell = 2$ mode induces the 
rapid increase of the gravitational 
waves from the matter motions, albeit its amplitudes being very small, and is 
 clearly understood from the quadrupole nature of gravitational radiations 
(compare the left and right panels of Figure \ref{saturation}). 
 
\begin{figure}[hbt]
\epsscale{1.1}
\plottwo{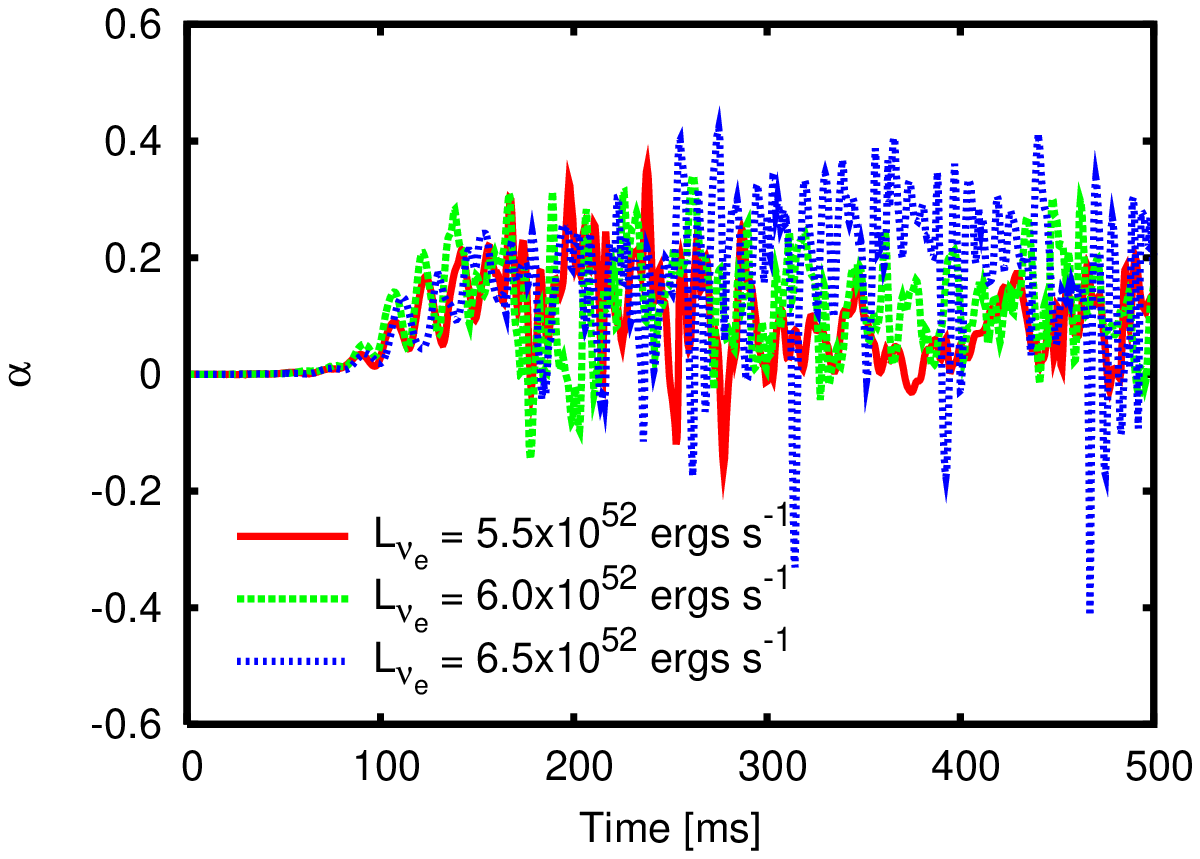}{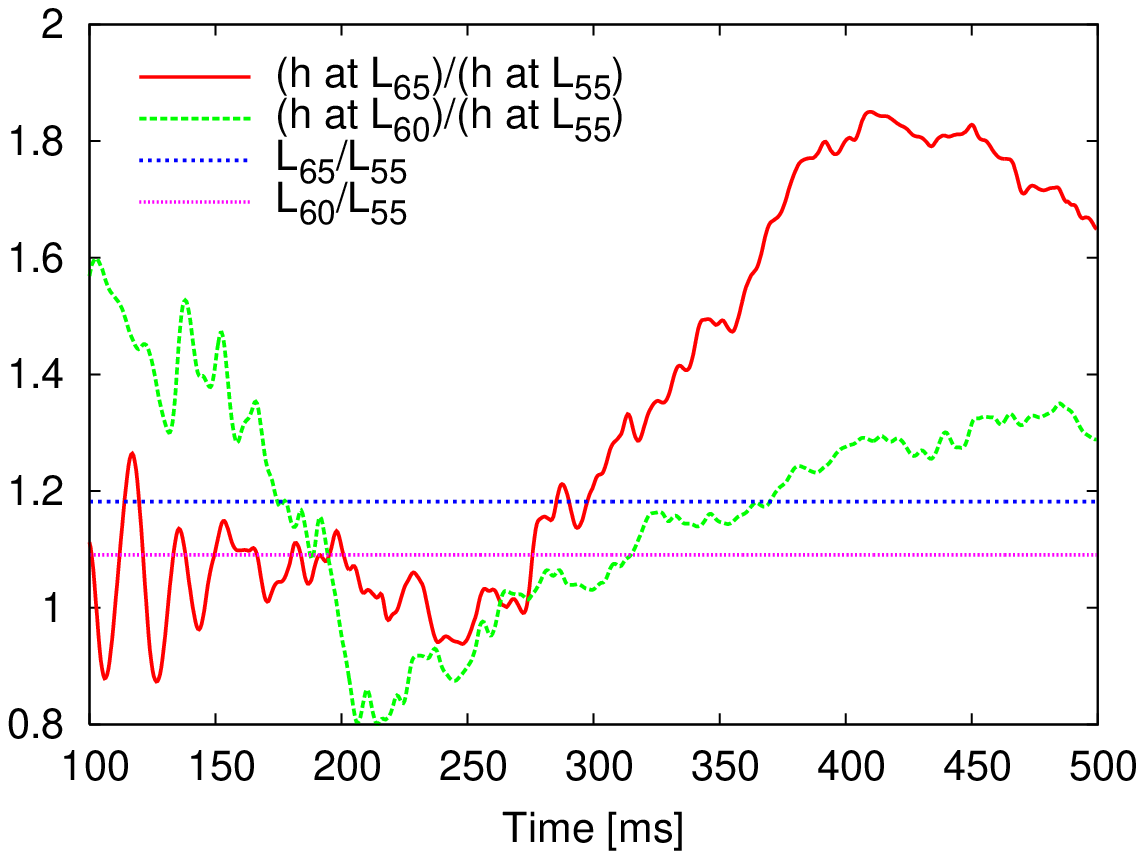}
\caption{Time evolution of the neutrino anisotropy parameter 
(see Eq. (\ref{anisop})) (left) and the ratios of the wave amplitudes normalized  by the case of the lowest input luminosity of 
$L_{\nu_e} = 5.5 \times 10^{52}~{\rm erg}~{\rm s}^{-1}~({\rm denoted~as}~ L_{55})$ (right). 
The horizontal lines in the right panel just represent the constant ratio of the input luminosity.  }
\label{alpha_figure}
\end{figure}

 The anisotropy parameter $\alpha(t)$ in Eq.(\ref{anisop}) also
 helps us to see the relation between 
the anisotropy of the neutrino radiation fields and the properties of the 
waveforms. The time evolution of $\alpha$ is presented in Figure 
\ref{alpha_figure}. In case of $L_{\nu_e} = 6.5 \times 10^{52}~{\rm 
erg}~{\rm s}^{-1}$,  $\alpha$ takes larger values after the saturation, 
leading to more greater wave amplitudes.
In the right panel of Figure \ref{alpha_figure}, the ratios of the wave 
amplitudes of $L_{\nu_e} = 6.5 \times 10^{52}~{\rm 
erg}~{\rm s}^{-1}$ and $L_{\nu_e} = 6.0 \times 10^{52}~{\rm 
erg}~{\rm s}^{-1}$ to $L_{\nu_e} = 5.5 \times 10^{52}~{\rm 
erg}~{\rm s}^{-1}$, are shown. Only near after the nonlinear phase 
sets in ($\sim 200$ ms), it is seen that 
the SASI-induced anisotropy of the neutrino emissions determines the
 waveforms, whose amplitudes monotonically grow with time.
As a reference, we draw the lines indicating the ratio of the input 
neutrino luminosities. It can be also seen that the increase of the 
amplitudes is greater than the one estimated only by the difference of
 the input luminosities after the onset of the nonlinear phase.

\begin{figure}[hbt]
\epsscale{1.1}
\plottwo{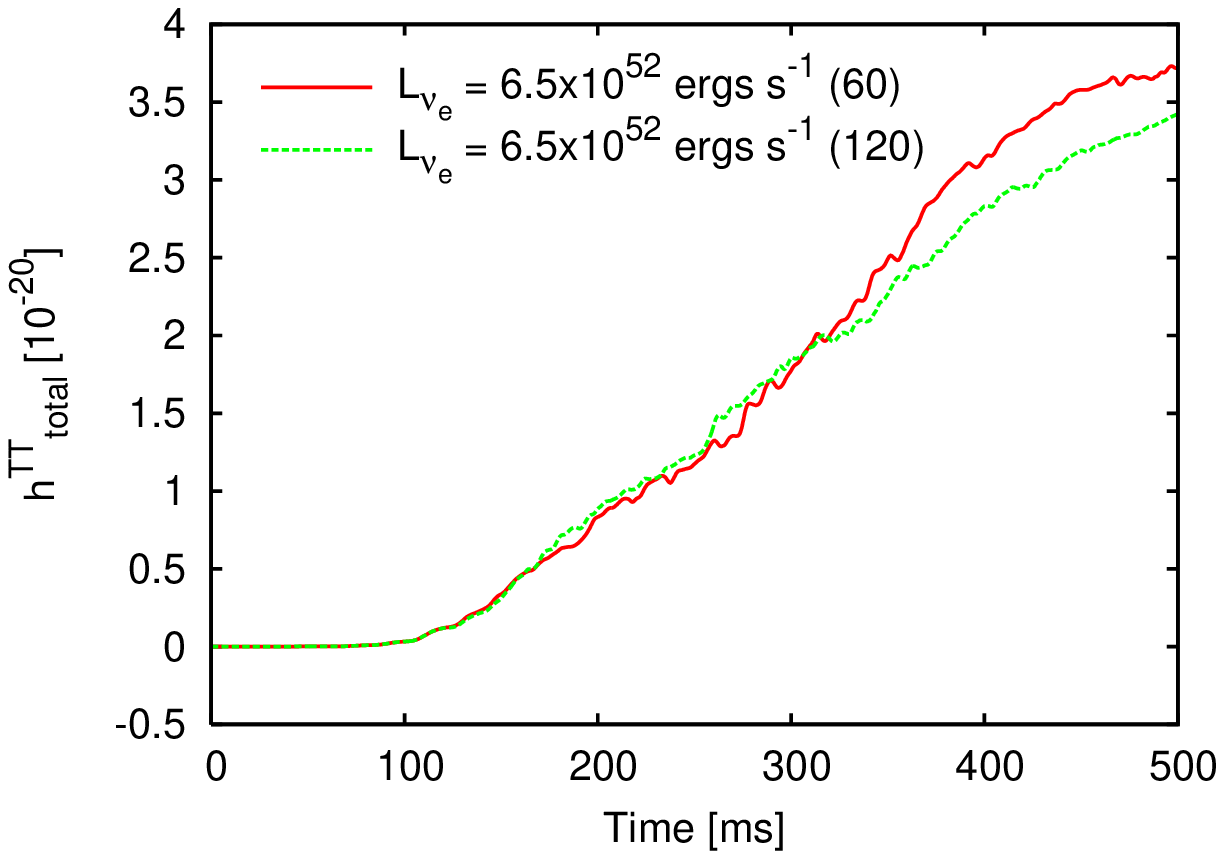}{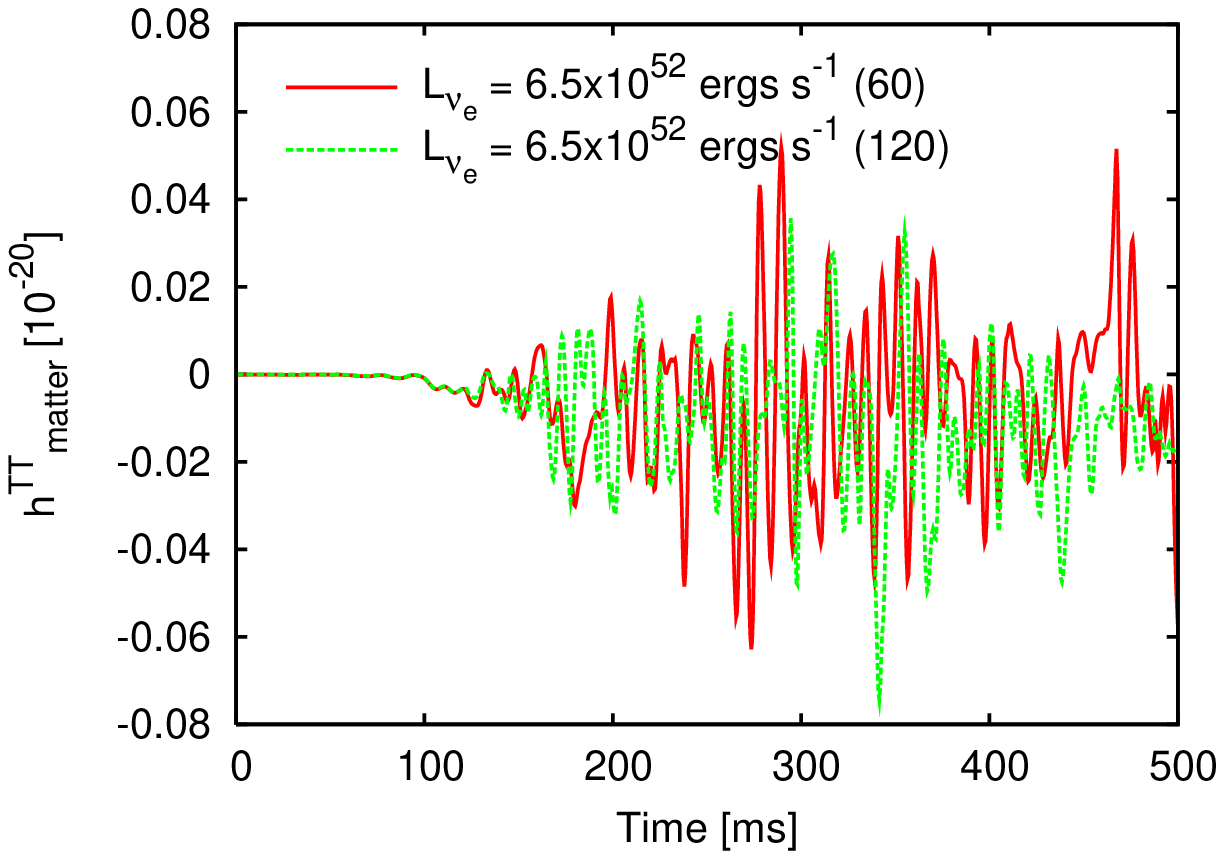}
\caption{Effects of numerical resolutions on the waveforms. 60, 120 in the panel represents the number of the angular mesh points in the computational domain. Left and right 
panel shows the waveforms from the sum of the anisotropic neutrino emissions 
and only from the matter 
motions, respectively. After the saturation occurs ($\sim 200$ ms), the difference 
of the resolution appears but not becomes larger than $\sim 20 \%$ in 
the total amplitudes (left). } 
\label{resol_figure}
\end{figure}

Finally, we have to discuss whether the above discussions are subject 
to change when the numerical resolution becomes more better. In Figure 
\ref{resol_figure}, we compare the waveforms when the angular resolution
 is doubled in comparison with the model computed so far. As easily guessed, 
 the difference appears after the SASI saturates ($\sim 200$ ms) and the 
non-linear phase sets in. In the lower resolution 
calculation, the anisotropy of the shock propagation tends to become
 more larger than the high resolution case, 
which results in the larger neutrino anisotropy, and thus 
leading to the larger wave amplitudes. The difference is not greater than 
$\sim 20 \%$, and thus the qualitative features discussed 
so far are found to be unchanged.    

\begin{figure}[hbt]
\epsscale{1.0}
\plottwo{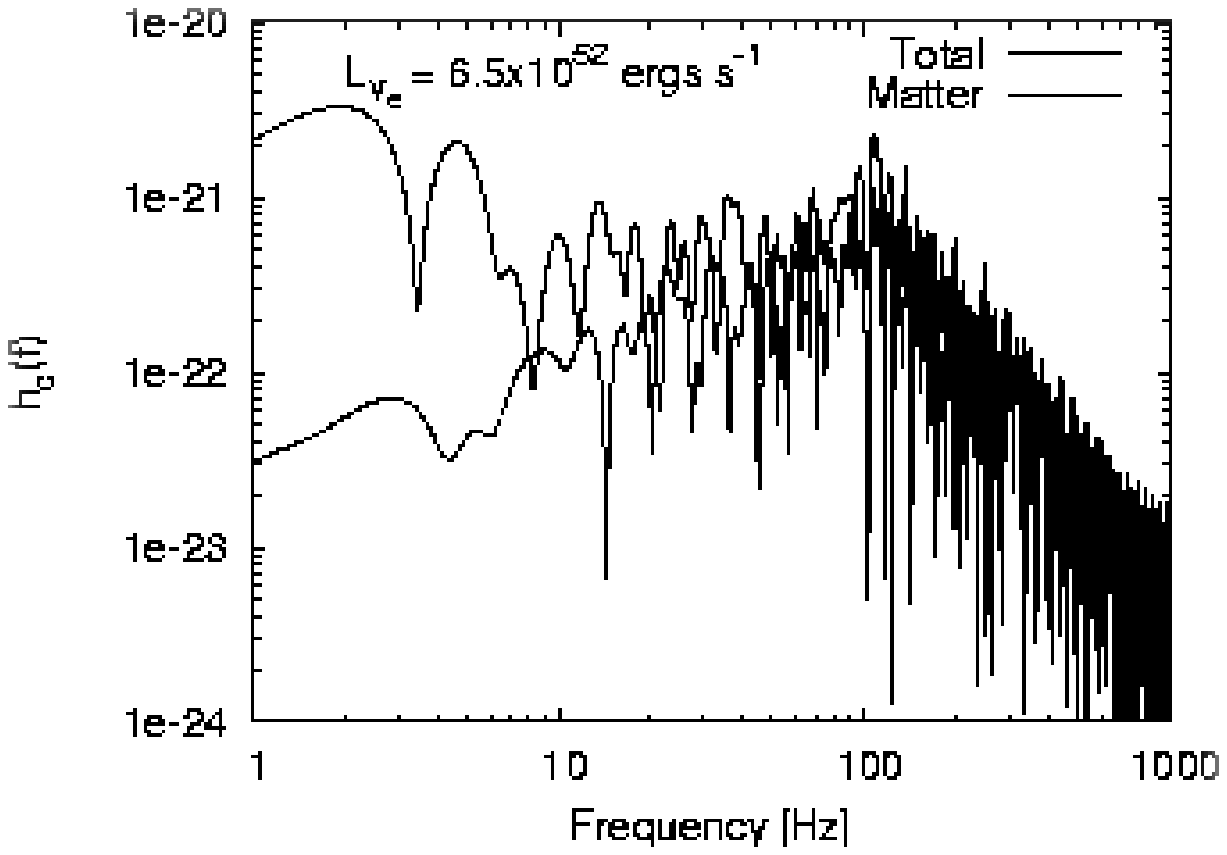}{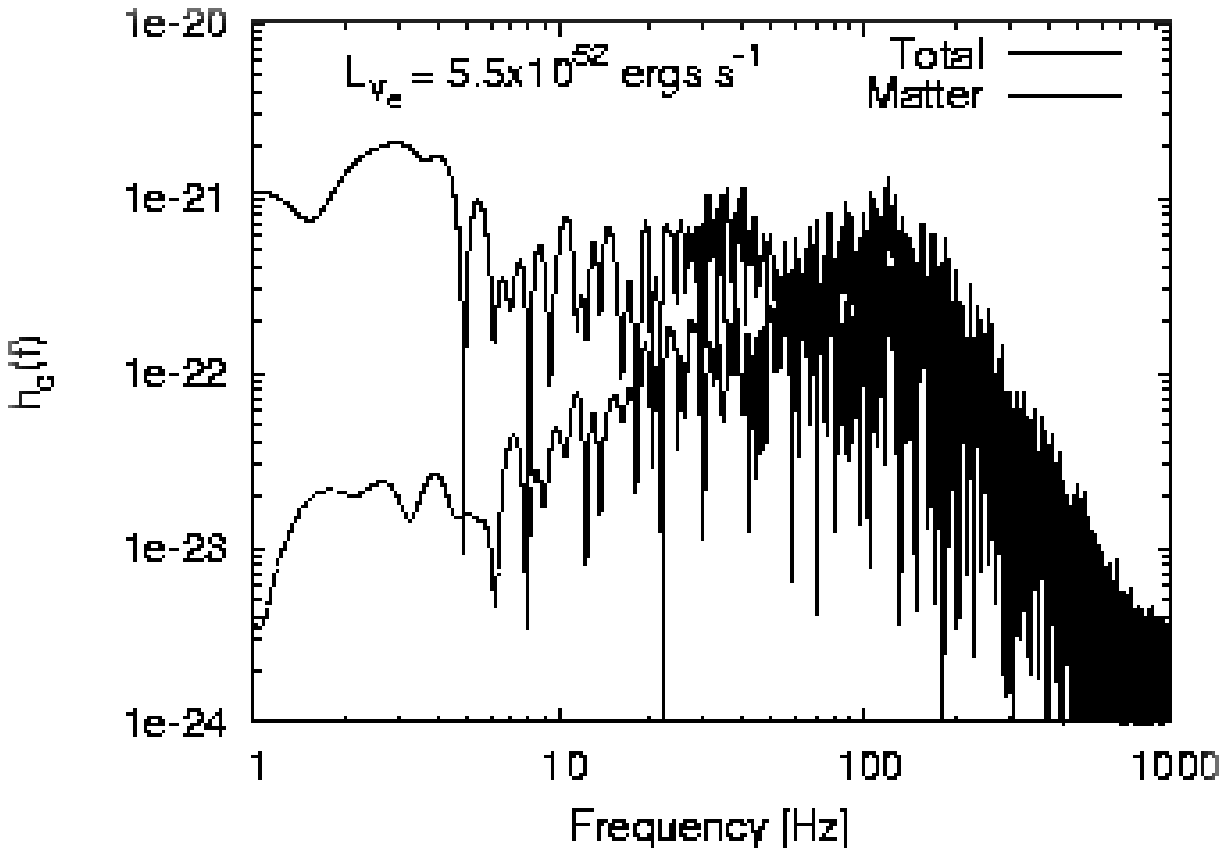}
\caption{Spectral distributions of the gravitational waves from the sum of 
anisotropic neutrino emissions and matter motions (thick line), and only from 
the matter motions (dotted line) for the case 
of $L_{\nu_e} = 6.5$ (left) and $5.5 \times 10^{52}~\rm{erg}~\rm{s}^{-1}$ 
(right), respectively. It can be seen the dominance of the neutrino-originated gravitational waves at the lower frequency of $\sim 100$ Hz. }
\label{spectrum}
\end{figure}

\subsection{Spectrum Analysis}
Now we move on to discuss the features of the waveforms by the 
 spectrum analysis. From Figure \ref{spectrum}, one can see the dominance 
of the neutrino-originated gravitational waves at lower frequency of 
$\sim 100$ Hz.
As seen in the left panel of Figure \ref{fig2}, the waveform from 
the neutrino shows a long-time variability in comparison with the rapidly 
varying ($O({\rm ms}))$ waveforms from matter motions, 
which is due to the local hydrodynamical instabilities. 
It is the memory effect of the neutrino-originated 
gravitational waves mentioned earlier that absorbs the 
rapid time variations of the neutrino anisotropy. 

\begin{figure}[hbt]
\epsscale{1.0}
\plotone{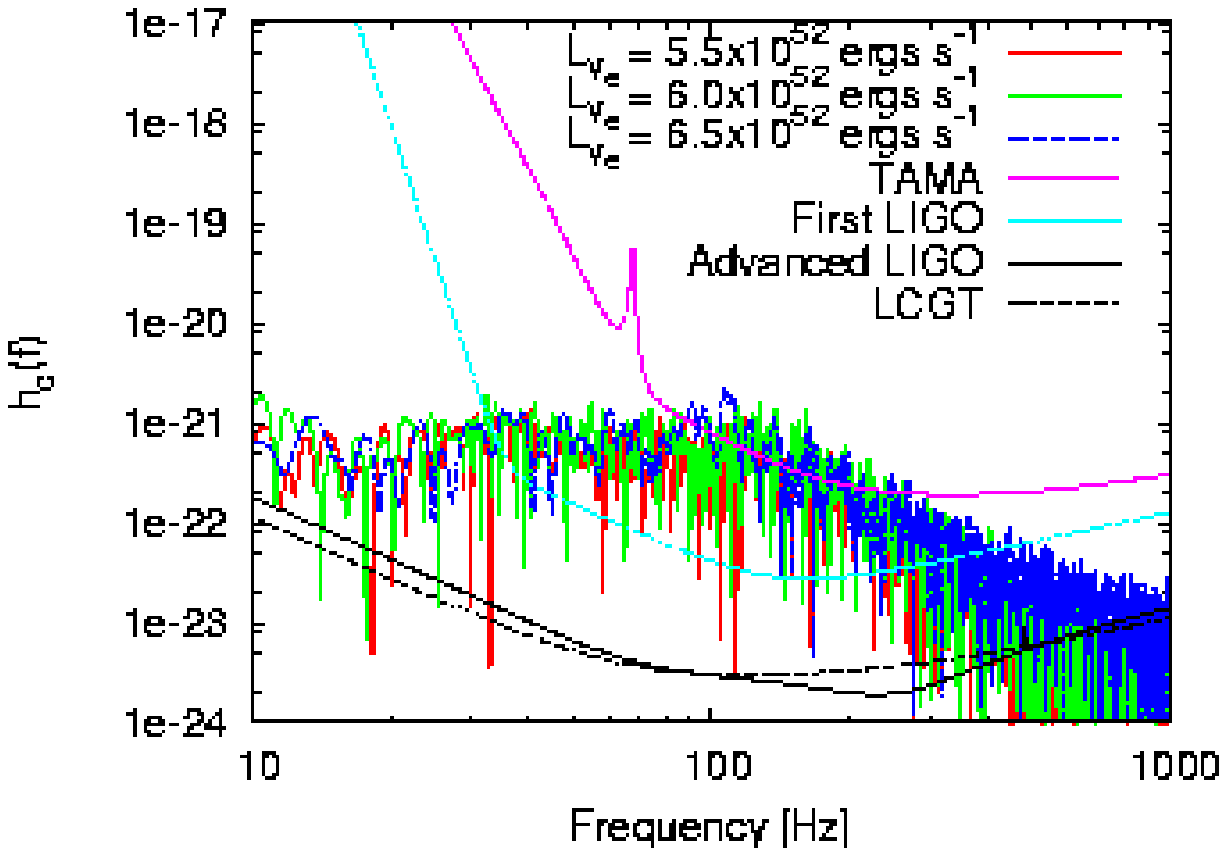}
\caption{Detection limits of TAMA \citep{tama}, first LIGO \citep{firstligo}, 
advanced LIGO \citep{advancedligo}, and Large-scale
 Cryogenic Gravitational wave Telescope (LCGT) \citep{lcgt} and 
 the expected gravitational wave spectra. The source is assumed to be located at the distance of 10 kpc.}
\label{all_spectrum}
\end{figure}

 To see clearly the dominance of the neutrino-originated 
gravitational wave than the matter-originated one at the lower frequency, 
we define $f_{\rm eq}$, the frequency below which the dominance occurs, 
and the corresponding gravitational wave amplitude, $h_{\rm eq}$ (see 
Table \ref{table1}). From the table, it is found
 that the typical frequency is smaller for the higher luminosity case.
In Figure \ref{all_spectrum}, the gravitational wave spectra are 
plotted with the sensitivity curves of the laser interferometers. 
It can be seen that the detection of the gravitational wave 
at the low frequency range becomes more promising thanks to 
the contributions from the neutrinos.
It can be seen that the gravitational waves from the neutrinos, which 
are dominated below $\sim 100$ Hz, seem marginally within the detection 
limits of the currently running detector of the first LIGO and the detection 
seems more feasible for the detectors in the next generation 
such as LCGT and the advanced LIGO if a supernova occurs in our galactic center.

\begin{figure}[hbt]
\epsscale{1.0}
\plotone{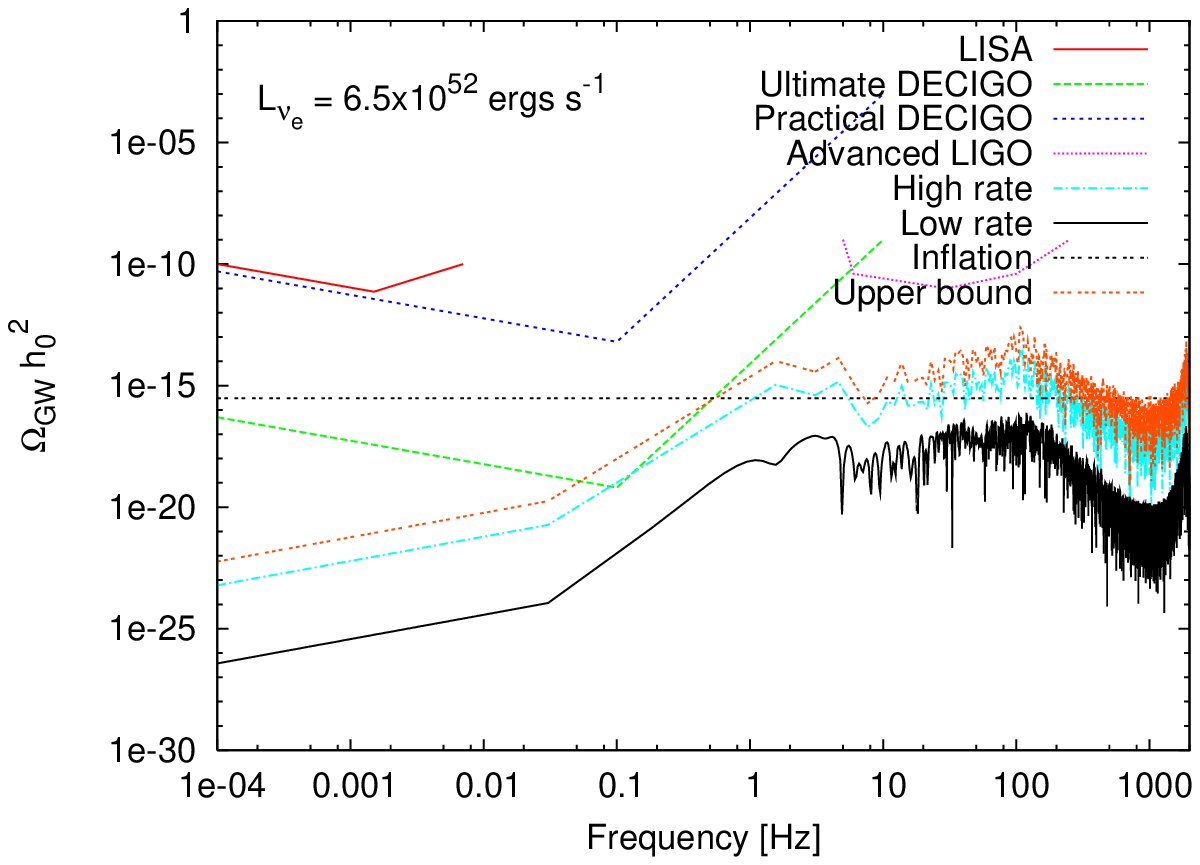}
\caption{Cosmological gravitational wave background for the highest
 luminosity case of 
$L_{\nu_e} = 6.5 \times 10^{52}~{\rm erg}~{\rm s}^{-1}$ 
, taking into account 
the possible enhancement due to the uncertainty of the cosmic star 
formation rate labeled by ``Low rate'' and ``High rate'', and the limited 
simulation time labeled by ``Upper bound'' (see text for details), 
with some indicated sensitivity curves for future missions of LISA, 
the LIGO in the third generation \citep{buonanno_tsi}, and practical/ultimate DECIGO \citep{seto}.
 The horizontal line
 indicates a maximum version of gravitational wave backgrounds produced
 during the slow-roll inflation assuming $T/S =0.3$ for 
the ratio of tensorial and scalar contributions to the cosmic microwave
 background radiation anisotropy \citep{turner_bg}, concreted in 
\citet{buonanno}.}    
\label{GWB}
\end{figure}

Finally, we are in a position to discuss the contribution of the 
gravitational waves mentioned so far to the background gravitational 
radiation. We calculate $\Omega_{\rm gw}$ from Eqs. (\ref{fourier2}) and 
 (\ref{los}) for the highest luminosity case of 
$L_{\nu} = 6.5 \times 10^{52}~{\rm erg}~{\rm s}^{-1}$. 
The resulting amplitude is presented in Figure \ref{GWB}. 
 For the frequency greater than $\sim 100$ Hz, the signal is dominated by the 
contribution from matter motions, while at lower 
frequencies the signal is dominated by the one from the neutrinos (compare 
Figure \ref{all_spectrum}). 
``Low rate'' and ``High rate''in the figure correspond to the cases of 
$\alpha=0, 2$ in Eq. (\ref{evol}), respectively, indicating the uncertainty 
of the core-collapse supernova rate as mentioned before. 
We terminated 
the simulation for the high luminosity case (see Table \ref{table1}),
 seeing the shock wave propagate out of the iron core. For the frequencies 
of $f \ll (2\pi \Delta t)^{-1} \sim~0.1~{\rm Hz}(\Delta t/ 500~{\rm ms})^{-1}$ obtained from our 
simulations, we extended the spectra using the zero 
frequency limit \citep{turner} as implemented in the study of \citet{buonanno}. 
``Upper bound'' denotes the most optimistic estimation. In the case of 
$L_{\nu_e} = 6.5 \times 10^{52}~{\rm erg}~{\rm s}^{-1}$, about 
30 $\%$ of the available gravitational binding energy of the neutron star
 of $3\times 10^{53}~{\rm erg}$ may be emitted 
during the simulation of $500$ ms followed here. 
From Figure \ref{fig2} showing the monotonic increase of the neutrino-originated amplitudes, we multiply the 
 amplitude for the case of ``High rate'' by an enhancement factor of 3 and consider 
it as an upper bound. Then from Figure \ref{GWB}, we may say that the 
background radiation considered here could be larger for the frequency 
above $\sim$ 1 Hz than the 
upper limit of the background radiation generated in the inflationary epoch 
(the horizontal line in the figure).
However, the frequency range is just outside the sensitivity of the 
proposed detectors, such as DECIGO \citep{seto}.

\section{Summary and Discussion \label{sec4}}
We presented the results of numerical experiments, in which we studied
how the asphericities induced by the growth of the standing accretion 
shock instability could produce the gravitational waveforms 
in the postbounce phase of core-collapse supernovae.
 To obtain the neutrino-driven explosions, we parametrized the neutrino fluxes 
emitted from the central protoneutron star and approximated the neutrino 
transfer by the light-bulb scheme.  
By doing the spectrum analysis of the 
waveforms, we investigated the detectability of the signals 
 from a single core-collapse supernova and the cosmological ones by
the ground-based and space-based laser interferometers, respectively.
Our main results can be summarized as follows.

1. The amplitudes of the gravitational waves from the anisotropic neutrino 
emissions are larger up to two orders of magnitudes than the ones from the 
matter motions during the SASI operations. It is found that 
the wave amplitudes from the neutrinos show the monotonic increase with time, 
regardless of the neutrino luminosities from the protoneutron star. We point 
out that this feature can be understood by the specific nature of SASI, which 
makes the deformation of the shock waves of $l=1,2$ modes dominant,
 leading to the enhanced neutrino emissions in the regions close 
to the symmetry axis. In fact, we show that the amplitudes become 
larger when the growth of the SASI enters the nonlinear phase, in which 
the deformation of 
 the shocks and the neutrino anisotropy become large.

2. From the spectrum analysis of the waveforms, 
we find that the amplitudes from the 
anisotropic neutrino emissions are dominant over the ones 
from the matter motions at frequency $\lesssim 100$ Hz. 
 The detection of such signals from a galactic supernova may be marginal for 
the currently running detector of the first LIGO and 
 promising for the detectors in the next generation such as LCGT and the 
advanced LIGO.
 
3. As for the background radiation, we indicate that the contribution of the 
gravitational signals considered here could be larger at 
frequency $\gtrsim$ 1 Hz than the primordial gravitational wave backgrounds
 generated in the inflationary epoch. Unfortunately, however, it is found that 
 this frequency range is just outside of the sensitivity 
of the proposed detectors, such as DECIGO.

We give a brief comparison with recently published models.
The monotonic increase with time in the wave amplitudes of the 
neutrino-originated gravitational waves is consistent with the model 
s15r of \citet{mueller}, in which the operation of the SASI was seen.
However, the amplitudes here are typically larger (up to 1 order 
of magnitude). This should be mainly because the 
neutrino luminosity here is taken to be higher than the one obtained in 
\citet{mueller}, which is about $ \sim 2 \times 10^{52}$ erg/s during the 
SASI operation. As for the total gravitational-wave 
energy emission, typical values of the 
computed models here ($\sim 10^{-10} M_{\odot} c^2$, see Table 1) 
are one order magnitude smaller than the one in \citet{mueller}. 
This should be 
owing to the excision of the protoneutron star, by which the contribution
 from the high frequency domain of the energy emissions are eliminated 
in this study. It should be noted that the larger 
oscillations of the protoneutron star in the postbounce phase \citep{burr_new} 
 and the resulting efficient gravitational emissions \citep{ott_new}
 cannot be treated in principle here.

 The simulation highlighted here is nothing but an idealized study for 
the physical understanding of relation between the asphericities 
induced by the SASI and the resulting gravitational waves. Remembering the 
caveats about the assumptions of the artificially constructed initial 
condition, the fixed accretion rate, the absorbing boundary condition, and the 
fixed neutrino luminosity and energies, it is by no means definitive at all.
Especially,  much better neutrino transfer is indispensable for more
reliable calculations of the neutrino-originated gravitational wave, which we 
only considered the radial transport.
One more major deficit is the axial symmetry assumed in the 
present two-dimensional (2D) simulations. In three-dimensional 
environments, the pronounced dominance 
of $l =1,2$ along the symmetry axis, which is a coordinate singularity in the 
2D computations, may become weaker, owing to the additional spatial degree 
of freedom in the azimuthal ($\phi$) direction. In the 3D case, 
 we think that the qualitative features of the plus mode waveform 
computed in the 2D case here will be unchanged, but quantitatively, we expect that the amplitudes become smaller 
owing to the reduced anisotropy along the symmetry axis. Thus the 
amplitudes calculated in this study could be an upper bound, in which the
maximal anisotropy of the shock waves and thus neutrino emissions outside 
the neutrino sphere could be achieved. 
Furthermore we think that it is interesting to investigate the properties of
 the cross mode gravitational waves, which are of genuine 3D origin and 
could be possibly produced from the transfer of the $l=1,2$ modes 
to some modes with nonvanishing $m$ in $Y_{lm}$.  
This study is a prelude to the forthcoming 3D simulations to clarify
 those aspects, which will be presented elsewhere.

\acknowledgements{K.K. expresses thanks to T. Takiwaki, Y. Suwa 
for useful discussions, T. Hiramatsu for informative discussions with respect 
to the gravitational-wave backgrounds, and K. Sato for continuing 
encouragements.
The numerical calculations were partially done on the
supercomputers in RIKEN and KEK (KEK supercomputer Projects No.02-87 and
No.03-92). This work was supported in part 
by 
Grants-in-Aid for the Scientific Research from the Ministry of
Education, Science and Culture of Japan (No.S14102004, No.14079202, 
No.14740166), and Grant-in-Aid for the 21st century COE program
``Holistic Research and Education Center for Physics of Self-organizing Systems''.}

\end{document}